\documentclass[12pt]{article}
\usepackage{amsmath, setspace}
\usepackage{graphicx}
\usepackage{lineno}

 \makeatletter       
 \renewcommand{\@biblabel}[1]{#1.}
 \makeatother
\begin{document}
\doublespacing
\noindent
{\LARGE\textbf{Shape recognition of volcanic ash by simple convolutional neural network}}
\\
\\
{\large D. Shoji$^{1*}$, R. Noguchi$^{2}$}\\
\\
1. \textit{Earth-Life Science Institute, Tokyo Institute of Technology, 2-12-1 Ookayama, Meguro-ku, Tokyo}\\
2. \textit{Volcanic Fluid Research Center, Tokyo Institute of Technology, 2-12-1 Ookayama, Meguro-ku, Tokyo}\\
$*$ Corresponding author: shoji@elsi.jp
\newpage
\section*{Abstract}
Shape analyses of tephra grains result in understanding eruption mechanism of volcanoes. However, we have to define and select parameter set such as convexity for the precise discrimination of tephra grains. Selection of the best parameter set for the recognition of tephra shapes is complicated. Actually, many shape parameters have been suggested. Recently, neural network has made a great success in the field of machine learning. Convolutional neural network can recognize the shape of images without human bias and shape parameters. We applied the simple convolutional neural network developed for the handwritten digits to the recognition of tephra shapes. The network was trained by Morphologi tephra images, and it can recognize the tephra shapes with approximately 90\% of accuracy. 

\newpage
\section{Introduction}
Shape and morphology of volcanic ash are affected by the process of eruption (e.g., Wohletz and Heiken, 1992; Gonnermann et al., 2015). Thus, we can understand the eruption style and the transport of plume of volcanoes by analyzing the shape of tephra grains. For the shape analysis of volcanic ash, precise discrimination of tephra grains becomes important. Grouping of tephra grains by shape and morphology has been performed qualitatively using pictures of ash (e.g., Heiken, 1986). However, the classification of tephra by hand depends on the people who recognize the grains. Thus, for the objective discussions on volcanic ash, quantitative analyses by shape parameters are conducted (Liu et al., 2015; Leibrandt and Le Pennec, 2015; Miwa et al., 2015; Fitch et al., 2017; Schmith et al., 2017). Although quantitative analysis is useful for the precise classification of the tephra shape independent of the bias of human, one problem is that we have to take care what kind of parameters should be used (e.g., Liu et al., 2015). So far, many parameters which define the tephra shape have been suggested (e.g., circularity or convexity). In addition to these parameters, fractal analyses also have been conducted (Maria and Carey, 2007; Miwa et al., 2015). Search for the appropriate parameter set is a complicated task to recognize the shape of ash grains.

Recently, in the field of machine learning, neural network has been developed. Neural network is composed of some layers which have nodes. Each node accepts weighted signals from the nodes of the previous layer, and sends signals to the next layer. By using training data set, the weights and biases are updated to minimize the error, which is calculated as a loss function. For the recognition of images, convolutional neural network has made a great success (e.g., Krizhevsky et al., 2012). At the convolutional layer, output signals are organized to conserve the pattern of initial images. Even a simple convolutional neural network with a few layers can recognize the images of handwritten digits called MNIST (LeCun, 1998) with approximately 99\% of accuracy (e.g., Saitoh, 2016) .  

Although training data set is required for learning of convolutional neural network, we do not have to determine the shape parameters to discriminate the shape of images (machine can recognize the shape without shape parameter). In addition, once the training is conducted, machine can recognize the shape of images independent of human bias. Regardless of these advantages, application of the convolutional neural network to tephra images has not been conducted.

In this work, using the simple network suggested for the recognition of MNIST data set, we apply the convolutional neural network to the determination of tephra shape. We categorize training and test data sets of tephra images into four types of the shape (blocky, vesicular, elongated and rounded), and calculate the accuracy of test data set. As shown below, the simple neural network for MNIST can recognize the tephra shape with high accuracy, and thus it can be a powerful tool for the morphological analyses of volcanic ash. 

\section{Method}
\subsection{Volcanic ash grain}
Images of volcanic ash are taken using an automated grain analyzer: Morphologi G3S$^{\mathrm{TM}}$ (Malvern Instrument$^{\mathrm{TM}}$). For analyses using images, Scanning Electron Microscope (SEM) can take clear images with detailed surface texture. However, it cannot take many images with a short time. On the other hand, lots of 2-D projected grain images can be taken easily by Morphologi. For the machine learning, many training data are required to achieve high accuracy. Thus, we use image data set by Morphologi in this work. The settings of our measurements are; under bottom lighting, using an automated sample dispersion system (SDU), $\times$5 magnification, 80 for the threshold for background separation. The size of each image is 50$\times$50 pixels. In this study, we used volcanic ash samples from three volcanoes: 1) Funabara scoria cone, Izu Peninsula, Japan (magmatic origin), 2) Nippana tuff cone, Miyakejima Island, Japan (phreatomagmatic origin), and 3) Myvatn rootless cones, Iceland (lava-water interaction origin; Thorarinsson, 1953). Each samples are sieved using screens after drying by heating in an oven.  In this study, we use grains at 2$\phi$ to 3$\phi$ fraction (125$\mu$m to 250$\mu$m).

Here, we defined four types of the tephra grain shape: blocky, vesicular, elongated and rounded (Fig. 1). Blocky grains have edges at nearly right angle and the shape is close to square (equant). Heiken and Wohletz (1992) show that blocky grains are typical in phreatomagmatic eruptions. Vesicular grains have inwardly convex and irregular shape. This type of grains might be formed by fragmentation by exsolution of volatiles in magma (Maria and Carey, 2007).  Elongated grains are relatively long and thin compared to blocky grains.  This type of grains might be elongated basaltic glass formed by lava fountain (e.g., Heiken and Wohletz, 1992) and/or elongated crystals (e.g., plagioclase). Rounded grains have smooth and rounded shape. Some of them have cauliflower-like outline. They might be formed by surface tension within a fluid droplet before they are chilled (Heiken and Wohletz, 1992) or magma-water interaction for cauliflower-type (e.g., White, 1996).

We selected images which have typical form of the four shapes. In our sample images, the number of blocky, vesicular, elongated and rounded images is 218, 193, 180 and 199, respectively. Of these images, we selected 50 images at each shape as the test images, which can be used for the calculations of network accuracy. The rest images are the training images for the learning of the network. Because the number of each image is not sufficient, we made the additional training images by rotating and turning the original images. Then, we selected 1000 pictures at each shape as the training data set. For the learning, we normalized the intensity of each pixel between 0 and 1 dividing by 255.

\subsection{Convolutional neural network}
For the learning of images, we set neural network as shown in Fig. \ref{fig2}, which is composed of one convolutional layer, one pooling layer and two fully connected layers (we modified the open source code published at https://github.com/oreilly-japan/deep-learning-from-scratch). Normalized intensities of tephra images are the initial input signals. The input signals are convoluted by the 50 convolution filters. The size and the stride width of the filter are set at 5 and 1, respectively. Thus the convoluted signals by one filter become 46$\times$46 block. Then, the size of output signals is reduced by the max pooling layer into 23$\times$23 block. After the pooling, the signals are sent to the fully connected layers. Thus, the first fully connected layer has 8450 nodes for accepting signals. The number of nodes of the next layer is set at 100. The last layer outputs probabilities of the four types of the shape. The network recognizes the shape of the input image from the highest probability. From 4000 training images (1000 images at each shape), 100 images are selected randomly as a batch, and weights and biases are updated 40 times, which is defined as "one epoch". Continuing this process, learning of the network is conducted. Using test images, we calculate the recognition accuracy of this network at each epoch.

\section{Results}
\begin{figure}[htbp]
\centering
\includegraphics[width=15cm]{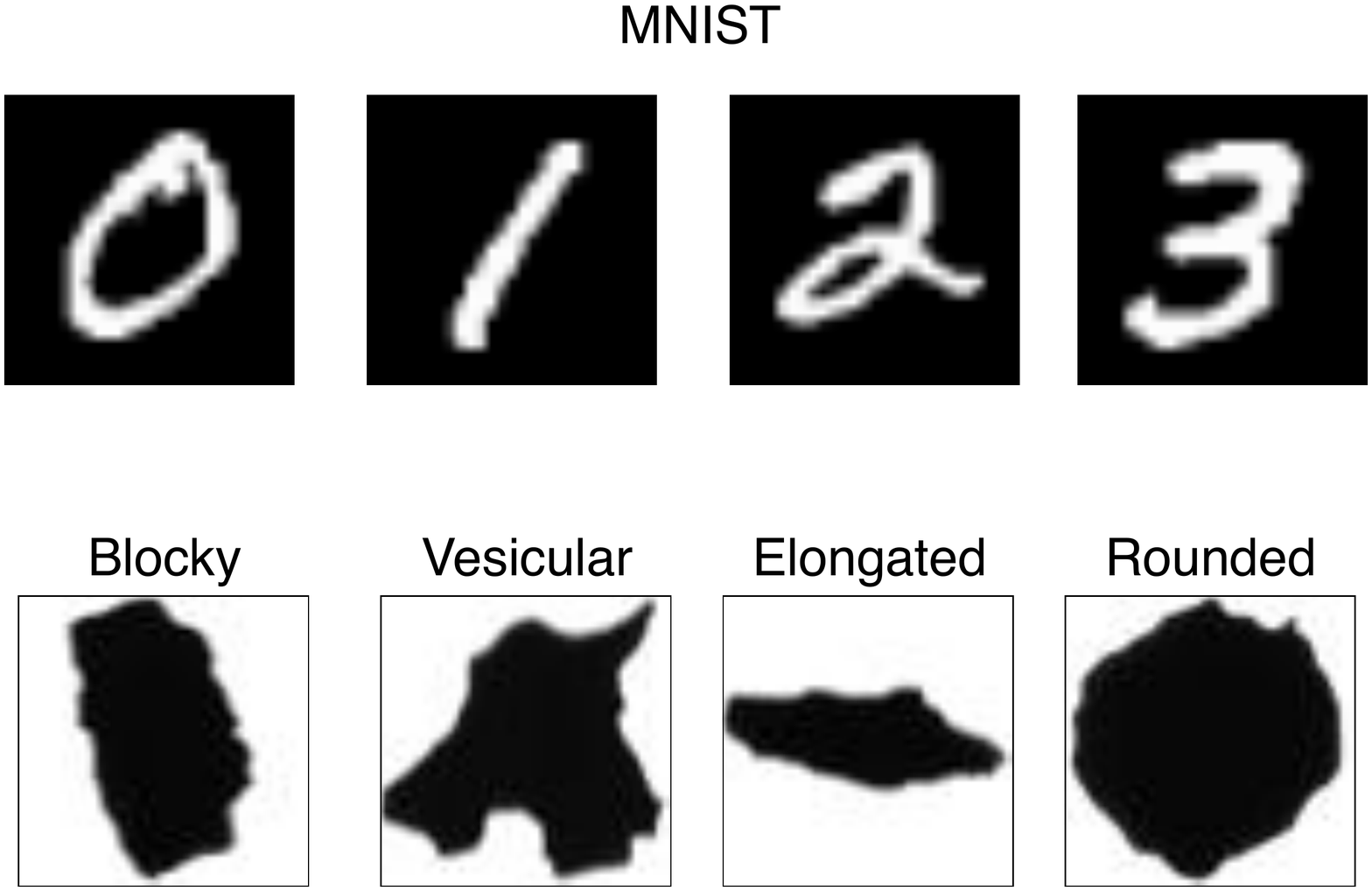}
\caption{Examples of MNIST images and tephra images by Morphologi. Both image sets are composed of gray scale pixels. While the size of MNIST data set is 28$\times$28 pixels, tephra images have 50$\times$50 pixels.}
\label{fig1}
\end{figure}

\begin{figure}[htbp]
\centering
\includegraphics[width=15cm] {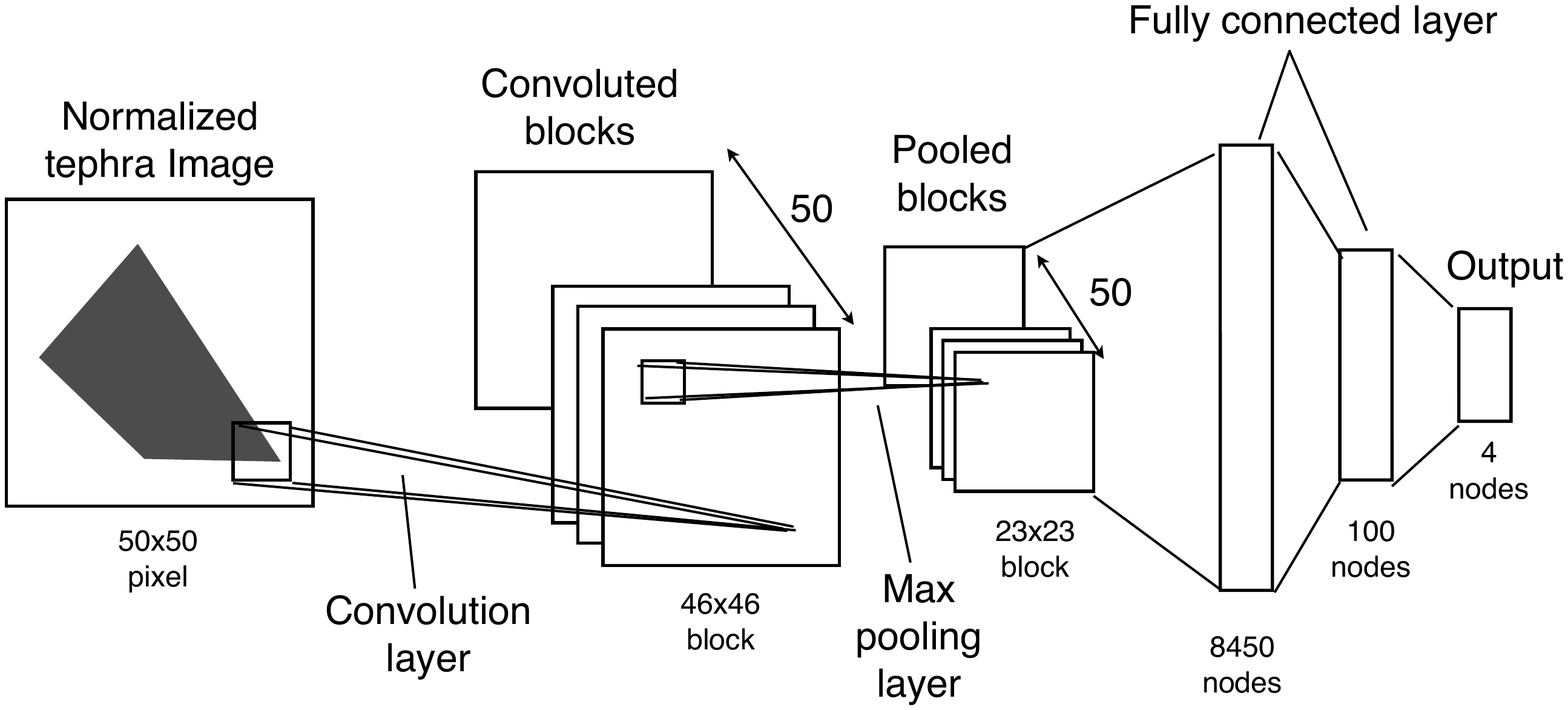}
\caption{Schematic view of the convolutional neural network. Intensities of tephra images are the input signals. Final layer outputs the probability of the shape of the input images.}
\label{fig2}
\end{figure}

\begin{figure}[htbp]
\centering
\includegraphics[width=12cm] {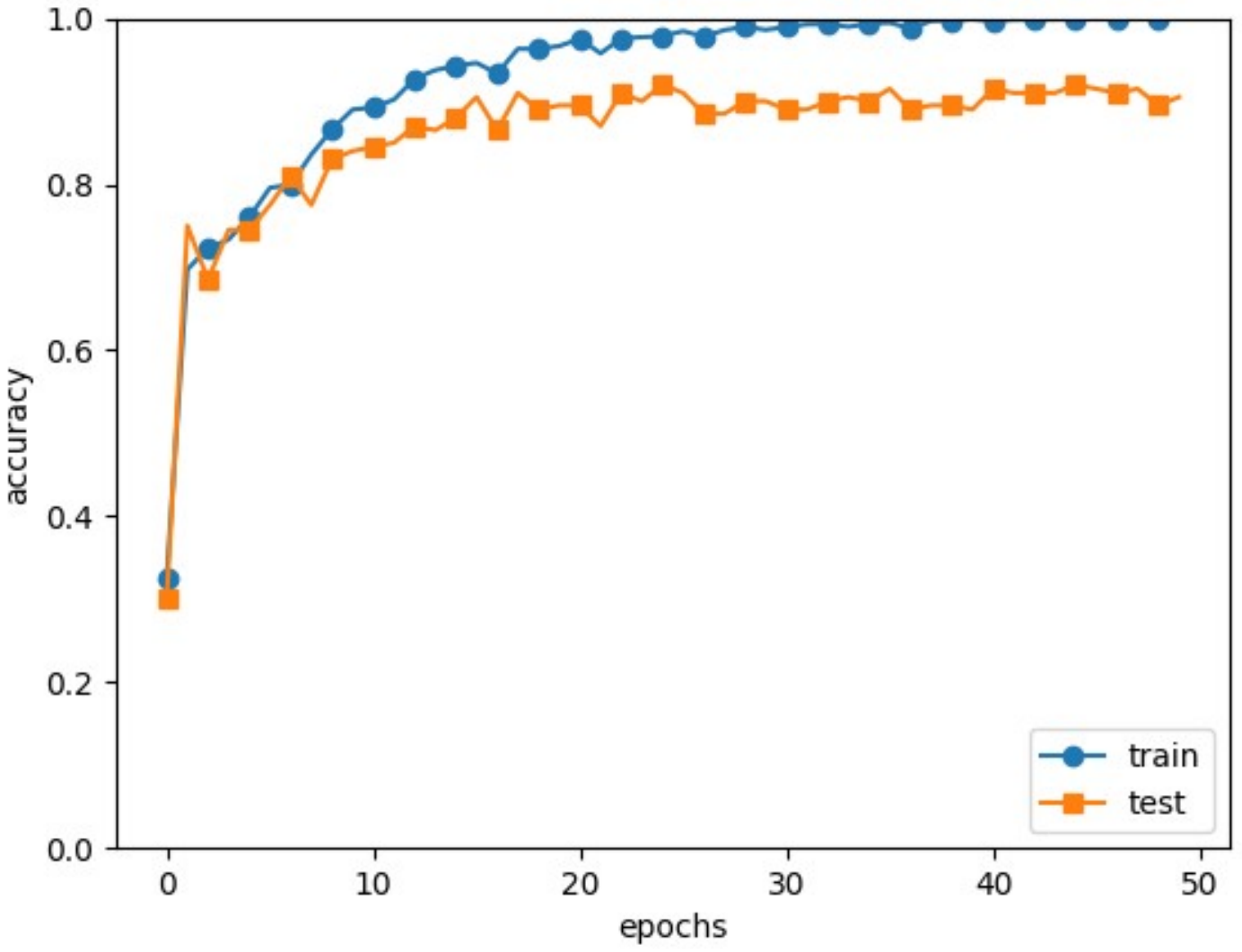}
\caption{Accuracies of the recognition of the training and the test images as a function of epoch.}
\label{fig3}
\end{figure}

Fig. \ref{fig3} shows the accuracies of the training images and the test images as a function of epoch. Both accuracies become stable at approximately 30 epochs. The network constructed in this work can recognize the shape of the test images with $\sim$91\% of accuracy. The tephra shapes by Morphologi images are composed of pixels of gray scale, which are similar to MNIST data set (Fig. \ref{fig1}). Thus, relatively high accuracy of recognition is attained by simple neural network. If a deeper network with many layers is used, accuracy may increase, which can be a future study. However, in the case of deep neural network, calculation time increases. 

\section{Conclusion}
Simple convolutional neutral network developed for the recognition of handwritten digits (MNIST) can be applied to the morphology analyses of volcanic ash. In the case of the determination of the four types of the tephra shape, our convolutional neural network can recognize the shape with approximately 90\% of accuracy. Thus, by making many tephra images by Mophologi and using convolutional neural network, shape analyses of tephra grains can be conducted without both human bias and shape parameters. Because morphological analysis is useful for the evaluation of the volcanic eruption style, convolutional neural network may become a tool for disaster preventions. We are preparing for the recognition of texture of grains as well as shape, and evaluating the effect of image number and size as a next step. 

\section*{Acknowledgements}
This work was supported by a JSPS Research Fellowship. We used Morphologi grain analyzer in National Institute of Advanced Industrial Science and Technology (AIST), Japan.  The samplings of volcanic ashes in Funabara and Myvatn were funded by Izu Paninsula Geopark Promotion Council and the Sasakawa Scientific Research Grant from The Japan Science Society for R.N (25-602) respectively.

\newpage

\end{document}